# Mn $L_{3,2}$ X-ray absorption and magnetic circular dichroism in ferromagnetic $Ga_{1-x}Mn_xP$

P.R. Stone, M.A. Scarpulla, R. Farshchi, I.D. Sharp, E.E. Haller, and O.D. Dubon[*]
*Department of Materials Science & Engineering, University of California, Berkeley, CA 94720 and Lawrence Berkeley National Laboratory, Berkeley, CA 94720*

K.M. Yu and J.W. Beeman
*Lawrence Berkeley National Laboratory, Berkeley, CA 94720*

E. Arenholz, J.D. Denlinger, and H. Ohldag[**]
*Advanced Light Source, Lawrence Berkeley National Laboratory, Berkeley, California 94720*

ABSTRACT

We have measured the X-ray absorption and X-ray magnetic circular dichroism (XMCD) at the Mn $L_{3,2}$ edges in ferromagnetic $Ga_{1-x}Mn_xP$ for $0.018 \leq x \leq 0.042$. Large XMCD asymmetries at the $L_3$ edge indicate significant spin-polarization of the density of states at the Fermi energy. The temperature dependence of the XMCD and moment per Mn of 2.67±0.45 $\mu_B$ calculated using sum rules are consistent with magnetometry values. The spectral shapes of the X-ray absorption and XMCD are nearly identical with those for $Ga_{1-x}Mn_xAs$ indicating that the hybridization of Mn $d$ and anion $p$ states is similar in the two materials.

[*]Email: oddubon@berkeley.edu
[**]Currently at Stanford Synchrotron Radiation Laboratory, Menlo Park, California, 94025

The discovery that conventional III-V semiconductors such as GaAs exhibit ferromagnetism when doped with a few atomic percent of Mn has led to unique possibilities for combined non-volatile information storage and processing [1]. Inter-Mn exchange is mediated in these ferromagnetic semiconductors by holes provided by substitutional manganese acceptors. In $Ga_{1-x}Mn_xAs$, the mediating holes occupy valence band-like states having some localized character. However, the nature of inter-Mn exchange across the Ga-Mn-pnictide series remains unresolved. We recently demonstrated the synthesis of a carrier-mediated ferromagnetic phase in $Ga_{1-x}Mn_xP$ using ion implantation and pulsed-laser melting (II-PLM) [2-4]. At 0.4 eV above the valence band edge, the Mn acceptor ground state in GaP [5] is significantly deeper than that in GaAs (0.11 eV) [6] leading to significantly more localized hole states. Despite this a ferromagnetic Curie temperature, $T_C$, above 60 K has been observed [3]. $Ga_{1-x}Mn_xP$ is thus an important medium for probing the interplay between electronic structure, localization and carrier-mediated exchange.

Characterization of the magnetic properties and spin-polarized band structure of $Ga_{1-x}Mn_xP$ is not only of scientific interest but also essential for assessing its potential for use as a source or detector of spin-polarized currents. Such information can be obtained from measurement of X-ray magnetic circular dichroism (XMCD) at the Mn $L_{3,2}$ absorption edges. Right or left circularly polarized light preferentially excites either spin down or up electrons from the spin-orbit split $p_{1/2}$ and $p_{3/2}$ Mn core states depending on the relative orientation of the spin and photon helicity vectors. These "spin rich" baths of electrons serve as a probe for the spin-polarized Mn 3$d$ contribution to the hole density of states at the Fermi energy ($E_F$) [7]. Further quantitative analysis can be carried out in conjunction with the X-ray absorption spectra of the Mn $L_{3,2}$ edge by applying the XMCD sum rules [8,9] to calculate the spin and orbital magnetic moments.

In this Letter we report X-ray absorption spectroscopy (XAS) and XMCD measurements of ferromagnetic $Ga_{1-x}Mn_xP$. XMCD spectra show asymmetries as high as 70% revealing strong spin polarization of mediating carriers and significant ferromagnetic coupling between Mn atoms. The XAS and XMCD spectra from $Ga_{1-x}Mn_xP$ are nearly identical in shape to those from $Ga_{1-x}Mn_xAs$ suggesting that the charge state and local environment of Mn atoms in the two materials are very similar.

Thin films of $Ga_{1-x}Mn_xP$ were synthesized by II-PLM [10]. GaP (001) wafers doped n-type with S ($1-6 \times 10^{16}$ /cm$^3$) were implanted with 50 keV Mn$^+$ to doses between $4.5 \times 10^{15}$ and $2.0 \times 10^{16}$ /cm$^2$. Samples measuring approximately 5 mm on a side were cleaved along <110> directions and irradiated in air with a single 0.44 J/cm$^2$ pulse from a KrF ($\lambda$=248 nm) excimer laser having FWHM of 18 ns. The beam was homogenized to a spatial uniformity of ±5% by a crossed-cylindrical lens homogenizer. Prolonged (24 hours) etching in concentrated HCl was used to remove a highly-twinned layer as well as any surface oxide phases. After processing the $Ga_{1-x}Mn_xP$ films were approximately 100 nm thick and characterized by a Mn concentration that reaches a maximum between 20 and 30 nm below the surface as determined by secondary ion mass spectrometry (SIMS). The concentrations of substitutional manganese, $Mn_{Ga}$, were determined by the combination of SIMS and ion beam analysis. We define x as the peak $Mn_{Ga}$ concentration by analogy to $Ga_{1-x}Mn_xAs$ as studies of this material formed by II-PLM have shown that its magnetic properties are governed by the film region with the maximum $Mn_{Ga}$ concentration [4].

DC magnetization was measured by SQUID magnetometry. The films studied had x=0.018, 0.029, 0.034, 0.038, and 0.042 with corresponding $T_C$s of 17, 35, 43, 52, and 60 K (±2K for all x), respectively. Room temperature XAS was performed at beamline 8.0 at the Advanced Light Source (ALS). Low-temperature XAS and XMCD measurements were carried out between 17 and 52 K in the vector magnetometer endstation at beamline 4.0 at the ALS in applied fields in the range of ±5.4 kOe [11]. Data were collected with the field and beam oriented 30º from the plane of the samples along a <110> in-plane direction with 90% circular polarization of the incident X-rays.

The main panel of Fig. 1 presents XAS data at the Mn $L_{3,2}$ edge taken in total electron yield (TEY) mode from a sample having x=0.042 before and after etching with HCl for 24 hours. The pre-edge absorption intensities have been normalized to unity and the main $Mn^{2+}$ absorption peak in the unetched sample has been calibrated to fall at an energy of 640.0 eV. The spectrum before HCl etching shows multiple sharp peaks characteristic of the unhybridized atomic Mn $d^5$ absorption spectrum similar to that seen for the Mn-rich surface oxide phases on $Ga_{1-x}Mn_xAs$, which obscured early XAS and XMCD measurements in $Ga_{1-x}Mn_xAs$ [12]. The atomic multiplets and higher energy peaks disappear after etching, resulting in the smoother Mn absorption spectrum characteristic of the orbital mixing between $Mn_{Ga}$ $d$ states and phosphorus $p$ states. Furthermore, the $L_3$ absorption peak shifts to a lower energy by approximately 0.5 eV, which is characteristic of the spectral change from atomic to hybridized Mn [13]. Corresponding effects can be seen at the oxygen $K$-edge, as demonstrated in the inset to Fig. 1. After etching the signal from oxygen is reduced by an order of magnitude demonstrating the removal of surface oxide phases. Therefore, the following XAS and XMCD signals obtained at low temperature arise from Mn in the GaP matrix and not from surface oxide phases.

The main panel of Fig. 2 presents Mn $L_{3,2}$ TEY XAS spectra taken at 17 K with the field and photon helicity parallel ($I^+$) and antiparallel ($I^-$) for a sample having x=0.034. The XMCD $\left((I^+ - I^-)/(I^+ + I^-)\right)$ spectrum is also shown. Linear backgrounds fit to the pre-edge region were subtracted from the raw data, and the data were normalized to the main $L_3$ peak. Strong XMCD is present at both the $L_3$ and $L_2$ edges indicating strong magnetization of Mn and a large spin polarization of states derived from Mn $d$ levels at $E_F$. Very similar spectra were obtained for all of the samples having different compositions as was reported for $Ga_{1-x}Mn_xAs$ [14].

While TEY mode probes depths of under 10 nm the total fluorescence yield (TFY) mode can probe depths on the order of tens of nanometers [15]. Thus, TFY XMCD is a better probe of the bulk magnetic properties of these films. The magnitude of the XMCD at 639.5 eV and 17 K is plotted as a function of Mn composition for both TEY and TFY modes in the inset of Fig. 2. When corrected for incident angle and photon polarization, the TFY data exhibit a maximum asymmetry value of around 0.70±0.04 in all samples except for the one having x=0.018. This is primarily because the $T_C$ of 18 K of this film is very close to the measurement temperature, and thus its magnetic order is disrupted by thermal fluctuations. The near-constant value of the TFY XMCD at the higher compositions indicates that at the measurement field and temperature the magnetization per Mn and spin polarization in the hole density of states is nearly constant [14]. The TEY data are generally lower than the TFY data, which is consistent with lower Mn concentration and magnetic coupling near the surface of the films. The only exception to this trend is the sample with x=0.018, which we attribute to a difference in the Mn incorporation and regrowth after the pulsed laser melting for low implant doses.

Spin, orbital, and total magnetic moments were calculated following the sum rule analysis method in Ref. 16 assuming a *d* electron count of 5.1 and a correction factor for the spin magnetic moment of 1.47 [14]. The nearly identical XAS and XMCD spectral shapes obtained for $Ga_{1-x}Mn_xP$ in this study and for $Ga_{1-x}Mn_xAs$ [14] support the use of these parameters. Only the TEY spectra were analyzed due to the significantly lower signal-to-noise ratio for TFY. The orbital and spin moments were calculated to be 0.12±0.01 $\mu_B$/Mn and 2.55±0.45 $\mu_B$/Mn, respectively, over the entire dose series resulting in a total magnetic moment of 2.67±0.45 $\mu_B$/Mn and a ratio of orbital to spin moments of 0.048± 0.007. This total magnetic moment is consistent with the values of 3-4 $\mu_B$/$Mn_{Ga}$ obtained by SQUID magnetometry [3]. The magnetic moment calculated from TEY data is probably an underestimate of the magnetization due to the lower concentration of Mn in the near-surface region.

Figure 3 compares the film magnetization measured by SQUID magnetometry, $M_{SQUID}$, and normalized XMCD TEY and TFY signals as a function of temperature for a sample with x=0.034. The magnetization data obtained with a measuring field of 50 Oe yield a film $T_C$ of 43± 2K. In a measuring field of 5 kOe, corresponding to the measuring field with which XMCD spectra were taken, sample magnetization (dashed line) persists at temperatures above $T_C$. This explains the stronger than expected spin polarization above $T_C$ when the sample in nominally paramagnetic. The identical temperature dependencies of magnetization and XMCD demonstrate that these two techniques are measuring the same magnetic phase even though the characteristic probe depths differ.

The similarity between the XAS and XMCD lineshapes for $Ga_{1-x}Mn_xP$ and those reported for $Ga_{1-x}Mn_xAs$ [12, 14] is remarkable. Because XAS and XMCD lineshapes are strongly influenced by the hybridization of the $t_2$-symmetric Mn $d$ orbitals with the neighboring anion $p$ orbitals, this suggests that the bonding and $p$-$d$ exchange between Mn and As or P in dilute alloys are substantially similar [17]. This experimentally confirms electronic structure calculations, which generally show very similar densities of states near $E_F$ for $Ga_{1-x}Mn_xAs$ and $Ga_{1-x}Mn_xP$ [18-20]. In dilute GaAs:Mn, the consensus view is that the isolated Mn acceptor has a quasi-hydrogenic $d^5$+bound hole ground state for which the hole states are composed primarily of As $p$ states, which form the valence band. In GaN:Mn the $d^4$ configuration is more stable, and hole states have a dominant $d$-like character [21]. In GaP:Mn conflicting results have been reported, and electronic structure calculations show a very small difference in energy between the $d^4$ and $d^5$ configurations [20, 22-23]. For the $Ga_{1-x}Mn_xP$ alloy, the hole states at $E_F$ are expected to have more localized Mn $d$ ($t_2$) character than for $Ga_{1-x}Mn_xAs$ but still contain substantial phosphorus $p$ character, leading to a stronger tendency for a non-zero, spin-polarized density of states separated by a gap from the valence band [18-20]. Thus, the scenario we have proposed whereby ferromagnetism in $Ga_{1-x}Mn_xP$ (x up to 0.042) is mediated by localized holes in a Mn-derived band [3] is consistent with the X-ray measurements reported herein.

This work is supported by the Director, Office of Science, Office of Basic Energy Sciences, Division of Materials Sciences and Engineering, of the U.S. Department of Energy under Contract No. DE-AC02-05CH11231. The authors thank K. W. Edmonds, Y. Suzuki, and R. Gronsky for valuable discussions. P.R.S. acknowledges support from a NDSEG Fellowship.

FIGURE CAPTIONS

**Figure 1** – Room temperature XAS spectra before and after etching in HCl for 24 hrs for a $Ga_{1-x}Mn_xP$ sample with x=0.042 at the Mn $L_{3,2}$ edge (main) and oxygen $K$ edge (inset).

**Figure 2** – (main) Mn $L_{3,2}$ TEY XAS spectra for magnetization and helicity parallel ($I^+$) and antiparallel ($I^-$) as well as the difference (XMCD) spectrum for a $Ga_{1-x}Mn_xP$ sample with x=0.034 measured at 17 K. (inset) TEY and TFY XMCD magnitude for the Mn $L_3$ peak at 639.5 eV versus x. The raw XMCD data have been corrected for non-unity X-ray polarization and incident angle by multiplying by 1.283.

**Figure 3** – Temperature dependence of magnetization ($M_{SQUID}$) measured at fields of 0.05 (solid line) and 5 kOe (dashed line) and normalized XMCD signals from TEY (squares) and TFY (circles) taken at 5 kOe for a $Ga_{1-x}Mn_xP$ sample with x=0.034. Sample magnetization was measured along an in-plane <110> direction.

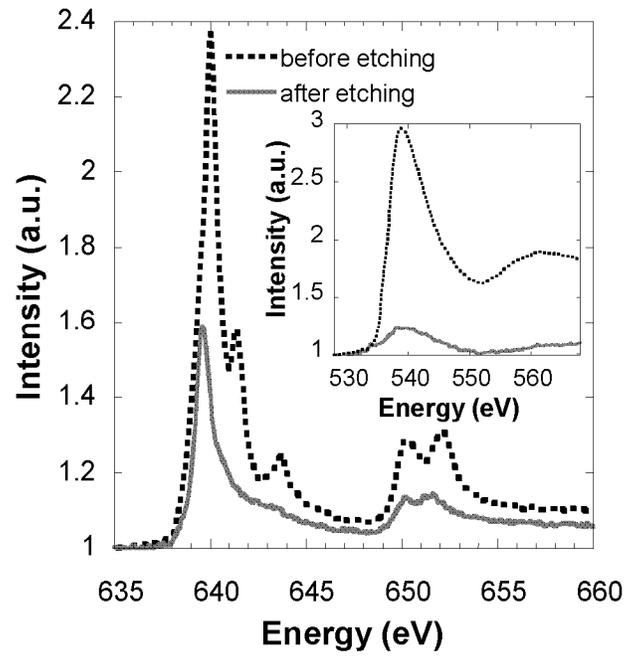

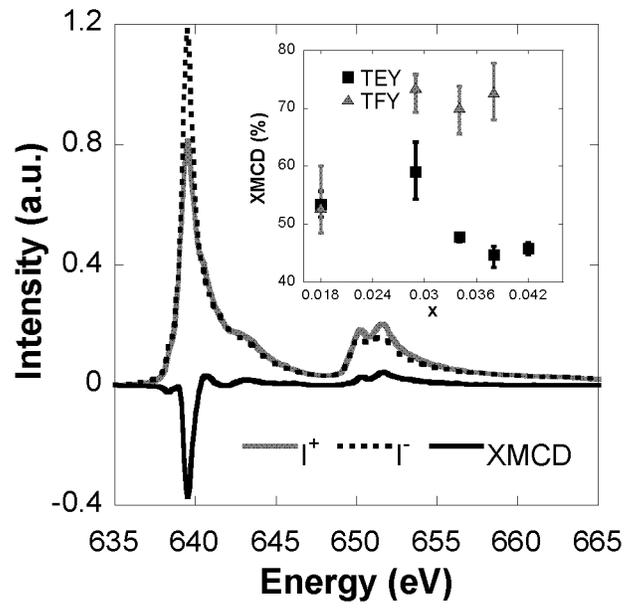

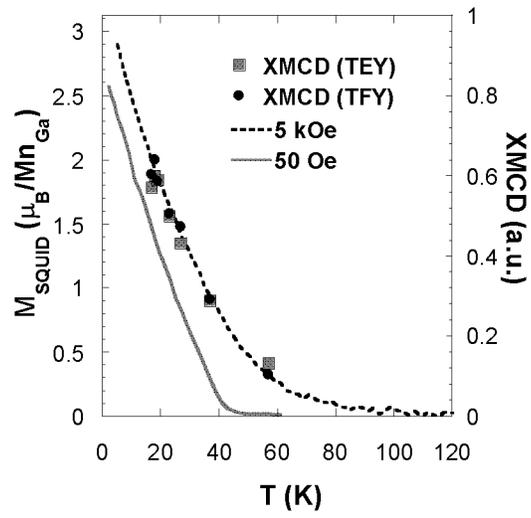